\documentclass[a4paper]{article}

\usepackage[english]{babel}
\usepackage[utf8x]{inputenc}
\usepackage[T1]{fontenc}

\usepackage[a4paper,top=3cm,bottom=2cm,left=3cm,right=3cm,marginparwidth=1.75cm]{geometry}

\usepackage{amsmath, amsthm, amssymb}
\usepackage{authblk}
\usepackage{mathrsfs}
\usepackage{graphicx}
\usepackage{subfigure}
\usepackage{float}
\usepackage{url}
\usepackage{bm}
\usepackage[colorinlistoftodos]{todonotes}
\usepackage[colorlinks=true, allcolors=blue]{hyperref}
\usepackage{booktabs}
\usepackage{algorithm}
\usepackage{algpseudocode}
\usepackage[numbers]{natbib}
\usepackage[titletoc]{appendix}
\usepackage{hyperref}
\usepackage{marvosym}

\numberwithin{equation}{subsection}

\title{\textbf{New methods derived from energy minimization problems for solving two dimensional discrete dislocation dynamics}}
\author[*]{Yuntong Huang $^{\href{mailto: yt_huang@whu.edu.cn}{\textrm{\Letter}}}$}
\author[*]{Shuyang Dai $^{\href{mailto: shuyang_dai@whu.edu.cn}{\textrm{\Letter}}}$}

\affil[*] {School of Mathematics and Statistics, Wuhan University, Wuhan 430072, China}

\begin{document}
\maketitle

\begin{abstract}
Dislocation dynamic is a typically gradient flow problem, and most of work solves it just as ODE, which means that the interacting energy of dislocations is ignored.
We take the interaction energy into account and use it to introduce new methods to speed up the simulation. 
The non-singular stress field theory is used to make sure that the interacting energy between dislocations is finite and computational, 
and using this the two dimensional discrete dislocation dynamics can be rewritten into optimal problems. 
Based on it, the new problems from 2D dislocation dynamics can be solved by conjugate gradient method and other optimal methods. 
We introduce several methods into dislocation dynamics from the energy point of view and some numerical experiments are presented to compare different
numerical methods, which show that the new methods are able to speed up relaxation procedures of dislocation dynamics. Those new approaches help to 
get the stable states of dislocations more quickly and speed up the simulations of dislocation dynamics.
\end{abstract}

\section{Introduction}

Predicting material properties, such as toughness of materials, is an important goal for materials scientists. Dislocations which are line defects in the crystal
structure are usually considered as the main reason for plastic deformation of crystal solids. Therefore, dislocation dynamics (DD) have been viewed as 
an important approach to simulate the behaviour of material and obtain crystal strength \cite{indenbomElasticStrainFields1992},
\cite{hirthTheoryDislocations1992}. 
Two-dimensional DD simulations usually focus on edge dislocations and each dislocation is modelled as a point in a certian plane, and
the 2D model is a further simplification of the 3D dislocation dynamics \cite{giessenDiscreteDislocationPlasticity1995}. In 3D DD, each dislocation is considered as several 
discrete segments, which are closer to real dislocations than points in 2D \cite{wangParallelAlgorithm3D2006}. The 3D model has been used to simulate some behaviours of crystal 
plasticity, but obviously the computational cost of it is very high. As a simplification, the 2D DD model is much easier to solve and is able to copy with
multiple slip systems of which all dislocations are straight and infinite. Dislocation sources and obstacles have also been introduced in 2D model. 
Now, the 2D model have attracted considerable attention and it has been successfully used to 
shed light on several aspects of materials, such as avalanche dynamics \cite{ovaskaQuenchedPinningCollective2015},
delivering a general picture of size effects \cite{derletMicroplasticityIntermittentDislocation2013} and so on.

In general, Runge-Kutta method(RK method) is widely used to solve the 2D model, which is an ordinary differetial equation (ODE) \cite{cartwrightDYNAMICSRUNGEKUTTA1992}. In order to speed up the simulation, 
people have proposed several approaches like an implicit time integration method \cite{peterffyEfficientImplicitTime2020}. But those measures definitely
still solve dislocation dynamics as ODE problems and there are a few physical properties of dislocations which have been ignored by previous research, such as 
the interacting energy. Drived from this idea, we solve 2D DD by solving the energy minimization problem of it. For minimization 
problems, there are many numerical methods for them like conjugate gradient methods\cite{nocedalNumericalOptimization2006}. However, it is obvious that our problems are non-linear and nonconvex, and 
some optimal approaches seem useless for our problems. \citet{wuInertialProximalGradient2021} proposed a 
Bergman proximal gradient method(BPG method), which is used in some nonconvex and complicated problems. We successfully apply it in our problems, 
and numerical tests demonstrated that the method speed up the simulations of 2D dislocation dynamics, compared to the normal RK method. However, we cannot tell
that the BPG method is less time consuming than the Runge-Kutta-Fehlberg method(RKF method) \cite{mathewsNumericalMethodsUsing2004}, which is very popular for dynamic systems. Then, we find that
Fletcher and Reeves conjugate gradient method(FRCG method)\cite{nocedalNumericalOptimization2006} can also be applied in our problems and the results of experiments show 
that this approach simulates more quickly than the RKF method. For the energy optimal problems, the energy dissipation property is important because it is related with the physical picture of dislocation dynamics.
However, the elastic energy might increase when we use RK method or RKF method, which solve models without considering the energy. 
If we use optimal methods with linear search methods, energy dissipation properties can be maintained naturally. What is more, numerical tests reveal that we can find the stable dislocation systems easier using 
FRCG method.

There are still few difficulties, like the singularity of stress field, which make it hard to solve 2D DD. In the development of 2D DD, much research 
concentrates on solving the problem of singularities intrinsic to the classical continuum theory of dislocations. Although the singular solutions 
of the continuum theory are simple, the energy and forces can be infinite and unsolvable. Therefore, some truncation schemes are applied to avoid 
this problem but some of them lack self-consistency of dislocation theory \cite{brownSelfstressDislocationsShape1964}, 
which means that the force related with local stress by Peach-Koehler formula should be the negative derivative of the interacting energy between dislocations. 
\Citeauthor{caiNonsingularContinuumTheory2006} has proposed a non-singular and self-consistency treatment, which ensures the energy and forces 
are finite and computational and makes it possible to consider dislocation dynamics as energy minimization problems. 
We succeed in changing DD into energy optimal problems for not only a single slip system but also a multiple slip system. 
It is also easy to consider periodic boundary conditions are usually used in the 2D model in our new problems.

This paper is organized as follows. Section 2 shows more detail about 2D dislocation dynamics and introduces types of boundary conditions in 2D DD. 
In Section 3, two optimal methods are presented for nonconvex and non-linear optimization, including BPG method and FRCG method. Moreover, those new methods can update time sizes
by linear search methods. In Section 4, the non-singular continuum theory of dislocations are discussed and our energy minimization problems are obtained 
from 2D DD. What is more, the BPG method is applied in those problems as an example. In Section 5, some numerical experiments are given to compare different 
approaches including BPG method, FRCG method and RKF method. Finally, we draw conclusion and discuss those numerical methods for discrete dislocation 
dynamics in Section 6.

\section{2D Dislocation Dynamics}
Two-dimensional(2D) discrete dislocation dynamics (DDD) simulations are a common procedure in dislocation research because they 
are easy to implement and light to compute but still show complex behaviour of crystal solids\cite{giessenDiscreteDislocationPlasticity1995},
\cite{derletMicroplasticityIntermittentDislocation2013}. In this model, only straight and parallel edge dislocations are considered, 
and it is enough to track positions of those dislocations on a plane perpendicular to the dislocation lines. 
Each dislocation \emph{i} is specified by three main parameters: position $\bm{x}_i=\left(x_i,y_i\right)$, 
Burgers vectors $\bm{b}_i=\left(b_{x,i},b_{y,i},b_{z,i}\right)$ and glide plane normal vector $\bm{n}_i=\left(n_{x,i},n_{y,i}\right)$. 
And in the following problems, for all dislocations \emph{i}, $b_{z,i}=0$.

The dislocations reside in a homogeneous linear elastic crystal. We will consider only three types of boundary 
conditions: (1) infinite in both \emph{x} and \emph{y}; (2) periodic in \emph{x} and infinite in \emph{y};
(3) periodic in both \emph{x} and \emph{y}. In last case, the simulation cell is a supercell with a square shape \cite{kuykendallConditionalConvergenceTwodimensional2013}.

\subsection{Case 1: Infinite in both \emph{x} and \emph{y}}
In this case, the dislocation lines are positioned parallel to the hidden \emph{z} axis and their Burgers Vector $\bm{b}=b\bm{u}_x$, 
are parallel to the \emph{x} axis which is also their glide direction. This means dislocations slip only along \emph{x} axis. The velocity of climbing of
dislocations is very small compared to gliding of dislocations, and then in the simutions climbing of dislocations is usually neglected.

The only relevant force per unit dislocation length may be calculated from the following expression between two edge dislocations 
(labelled \emph{i} and \emph{j}) on \emph{x} axis \cite{hirthTheoryDislocations1992}
\begin{equation}
    f_{x,ij}=\frac{\mu b_{x,i}b_{x,j}}{2\pi\left(1-\nu\right)}\frac{\Delta x\left(\Delta x^2-\Delta y^2\right)}{\left(\Delta x^2+\Delta y^2\right)^2}
    \label{1}
\end{equation}
where $\mu$ is the isotropic shear modulus and $\nu$ is Possion's ratio of the isotropic elastic medium, $b_{x,i}(b_{x,j})$ is the
Burgers vector in the \emph{x} direction of the \emph{ith}\emph{(jth)}, and $\left(\Delta x,\Delta y\right)$ is the two-dimensional 
vector defining the dislocations' spatial separation.

The equation of motion along the \emph{x} axis for the \emph{ith} dislocation is then given by
\begin{equation}
    \frac{dx_i}{dt}=\frac{F_{x,i}}{B}
    \label{2}
\end{equation}
where \emph{B} is the damping coefficient. In this equation. $F_{x,i}$ is the force per unit dislocation length acting on the dislocation.
\begin{equation}
    F_{x,i}=\sum_{j \neq i}f_{x,ij}+\sigma_{12}b_{x,i}
    \label{3}
\end{equation}
$\sigma_{12}$ is the applied shear stress.

In this model, $\frac{dy_i}{dt}=0$ for each dislocation \emph{i} because climbing of dislocations is not considered. Here is one of the simplest cases of dislocation dynamics. 
Generally speaking, the Runge-Kutta method would be used to solve equation \ref{2}, which is efficient for ODEs.

\subsection{Case 2: Periodic in \emph{x} and Infinite in \emph{y}}
For the dislocation systems with periodical conditions only in \emph{x} direction, the equation of motion along the \emph{x} axis for the \emph{ith} dislocation is the same as equation \ref{2}. 
However, the correct treatment of periodicity involves the simulation of all dislocation image contributions to the force per unit dislocation length on a given dislocation. 
So, considered one-dimensional periodicity, an exact solution to such a summation is tractable, and is given by\cite{kuykendallConditionalConvergenceTwodimensional2013}
\begin{equation}
\begin{aligned}
    f_{x,ij}&=\frac{\mu b_{x,i}b_{x,j}}{2\left(1-\nu\right)L}\frac{\sin{\Delta x_p}\left(\cosh{\Delta y_p}-\cos{\Delta x_p}-\Delta y_p\sinh{\Delta y_p}\right)}{\left(\cosh{\Delta y_p}-\cos{\Delta x_p}\right)^2}\\
    \frac{dx_i}{dt} &= \frac{1}{B}\left(\sum_{j \neq i}f_{x,ij}+\sigma_{12}b_{x,i}\right)\\
    \label{4}
\end{aligned}
\end{equation}
where $\Delta x_p=\frac{2\pi\left(x_i-x_j\right)}{L}$ and $\Delta y_p=\frac{2\pi\left(y_i-y_j\right)}{L}$, 
\emph{B} is the damping coefficient and \emph{L} is the length of the simulation box among \emph{x} axis. $\sigma_{12}$ is the external
applied stress.

The periodical condition in \emph{x} direction means
\begin{equation}
    x_i = x_i\, mod\, L + \left(x_i<0\right)?L:0\;\;\;\forall i=1,2,\cdot,N
    \label{5}
\end{equation}
where $mod$ is modular arithmetic.

It is easy to get similar results when it comes to be periodic in \emph{y} and infinite in \emph{x}, so we do not talk about that problem any longer.

\subsection{Case 3: Periodic in both \emph{x} and \emph{y}}
For dislocation systems which are applied with periodical conditions in both directions, it is easier to consider adjacent periods rather than the infinite sums, 
which is actually conditionallt convergent. In practice, truncating infinite sum is available and the square scheme (as the following Figure \ref{fig0}) is used in order to make the simulation be performed numerically 
\cite{kuykendallConditionalConvergenceTwodimensional2013}. We can include contributions from all image cells within a certain distance and increase this distance
until the value converges to the desired accuracy. Although we notice that the numerical sum depends on the order of summation, which means that this sum is 
conditionally convergent, we still use this measure to simulate dislocation behaviours under doubly PBCs because the square scheme is easy and effective to compute.

\begin{figure}[H]
    \centering
    \includegraphics[width=0.5\textwidth]{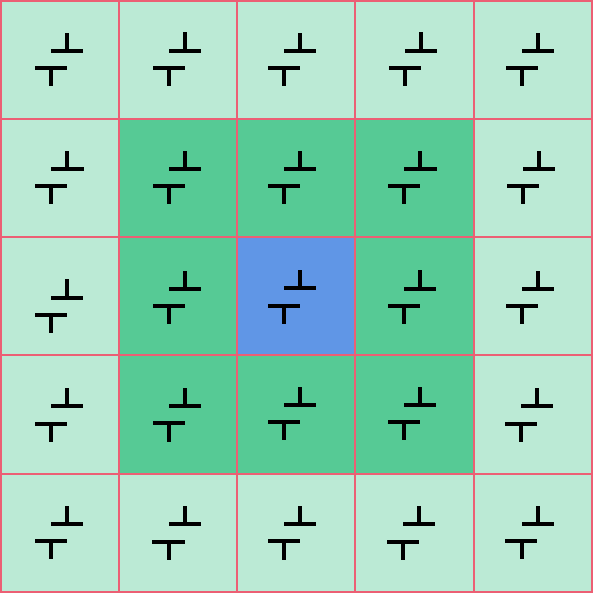}
    \caption{The truncation scheme: the blue cell is the primary cell, and the darker and the lighter shows smaller and larger cut-off regions included in the sum, respectively.}
    \label{fig0}
\end{figure}

The mobility equation of dislocation \emph{i} is:
\begin{equation}
    \Dot{\bm{x}_i}=\frac{1}{B}\left(\sum_{j\neq i} \bm{F}_{ij}+\bm{F}_{external}\right)
    \label{6}
\end{equation}
where $\bm{F}_{external}=\left(\bm{b_i}\cdot \bm{\sigma}_{external}\right)\times\zeta$ is the external applied stress field, $\bm{x}_i=\left(x_i, y_i\right)$ is the position of dislocation \emph{i}.
$\zeta$ is the dislocation line vector which is $\left(0,0,1\right)$ for all dislocations in systems.

And $\bm{F}_{ij}$ is given by Peach-Koehler equation\cite{lubardaDislocationBurgersVector2019}:
\begin{equation}
    \bm{F}_{ij}=\left(\bm{b_i}\cdot \bm{\sigma}^p\right)\times\zeta
    \label{7}
\end{equation}
where $\bm{\sigma}^p=\sum_{m,n=-N_c}^{N_c}\bm{\sigma}^{cell}\left(x_j,y_j,m,n\right)$ is the stress field under periodic conditions,
$N_c$ is the number of truncation terms. $\bm{\sigma}^{cell}$ reads:
\begin{equation*}
    \bm{\sigma}^{cell}\left(x_j,y_j,m,n\right)=\bm{\sigma}\left(x_i-x_j-m\emph{L},y_i-y_j-n\emph{L}\right)
\end{equation*}
The periodical conditions in \emph{x} and \emph{y} direction means
\begin{equation}
    \bm{x}_i= \bm{x}_i\, mod\, L + \left(\bm{x}_i<0\right)?L:0\;\;\;\forall i=1,2,\cdot,N
    \label{8}
\end{equation}
where \emph{L} is length of the simulation box, and $mod$ is modular arithmetic.

Here are three types of dislocation dynamic problems which we will consider and solve using optimal approaches in this paper.

\section{Methods For Optimization problem}
Dislocation dynamics usually are solved by Runge-Kutta methods, because those problems are usually viewed as ODE systems.
In fact, dislocation dynamics find the stationary states of systems of which elastic energy is a minima from a physical point of view. So, we introduce 
several optimal methods into dislocation dynamics, such as Bergman proximal gradient method and conjugate gradient method.
\subsection{The BPG Method}
The Bergman proximal gradient (BPG) method has been successfully applied in a few fields, including image processing and machine learning, and previous research has 
proved that it is efficient when it comes to solving nonconvex minimization problems \cite{bolteProximalAlternatingLinearized2014}, 
\cite{liuDualBregmanProximal2021}, \cite{mukkamalaConvexConcaveBacktrackingInertial2020}. For each iteration, this approach can update 
the positions of dislocations by solving a easy minimization problem.

Considering the minimization problem which has the following form \cite{jiangEfficientNumericalMethods2020}
\begin{equation}
    \mathop{min}_x E\left(\bm{x}\right)=f\left(\bm{x}\right)+g\left(\bm{x}\right)
    \label{9}
\end{equation}
where $f\in C^2$ is proper but nonconvex and $g$ is proper and convex. 
Let the domain of $E$ be $dom\mathop{E}=\left\{\bm{x}|E\left(\bm{x}\right)<+\infty\right\}$.

And we make the following assumptions: $E$ is bounded below, and for any $\bm{x^0}\in dom\mathop{E}$, the sublevel set 
$\left\{\bm{x}|E\left(\bm{x}\right)\leq E\left(\bm{x^0}\right)\right\}$ is compact.

Let \emph{h} be a strongly convex function such that $dom\mathop{h}\subseteq dom\mathop{f}$ and 
$dom\mathop{g}\cap intdom\mathop{h}\neq\emptyset$. Then according to it, the \emph{Bergman divergence} can be defined as
\begin{equation}
    D_h\left(\bm{x},\bm{y}\right)=h\left(\bm{x}\right)-h\left(\bm{y}\right)-\left\langle\nabla h\left(\bm{y}\right),\bm{x}-\bm{y}\right\rangle \;\forall \left(\bm{x},\bm{y}\right)\in dom\mathop{h}\times intdom\mathop{h}
    \label{10}
\end{equation}
It is easy to find that $D_h\geq 0$ and $D_h\left(\bm{x}, \bm{y}\right)=0$ if and only if $\bm{x}=\bm{y}$.

Bregman distance-based proximal methods have been proposed and applied for solving many nonconvex problems. Now, we would use it to solve problem \ref{9}. Basically, 
given the current estimation $\bm{x}^k\in intdom
\mathop{h}$ and step size $\alpha^k>0$, it updates $\bm{x}^{k+1}$ via
\begin{equation}
    \bm{x}^{k+1} = \mathop{argmin}_{\bm{x}}\left\{g\left(\bm{x}\right)+\left\langle\bm{x}-\bm{x}^k,\nabla f\left(\bm{x}^k\right)\right\rangle+\frac{1}{\alpha^k}D_h\left(\bm{x},\bm{x^k}\right)\right\}
    \label{11}
\end{equation}
$\alpha^k$ is the step size, which can be chosen by linear search.

$\bm{Step\, size\, estimation}$. In each step, $\alpha_k$ can be chosen by the backtracking linear search method and it would be initialized by the BB step estimation \cite{barzilaiTwoPointStepSize1988},
\begin{equation}
    \alpha^k=\frac{\left\langle\bm{s}^k,\bm{s}^k\right\rangle}{\left\langle\bm{s}^k,\bm{v}^k\right\rangle}\, or\, \frac{\left\langle\bm{v}^k,\bm{s}^k\right\rangle}{\left\langle\bm{v}^k,\bm{v}^k\right\rangle}
    \label{12}
\end{equation}
where $\bm{s}^k=\bm{x^k}-\bm{x^{k-1}}$ and $\bm{v}^k=\nabla f\left(\bm{x^k}\right)-\nabla f\left(\bm{x^{k-1}}\right)$. Let $\gamma >0$ be a small constant and $\bm{x}^{k+1}$ be obtained from \ref{11}; 
then the step size $\alpha_k$ should be chosen to hold the following inequality:
\begin{equation}
    E\left(\bm{x}^k\right)-E\left(\bm{x}^{k+1}\right)\geq\gamma\Vert\bm{x}^k-\bm{x}^{k+1}\Vert^2
    \label{13}
\end{equation}

More details are presented in the following Algorithm 


\Citeauthor{mukkamalaGlobalConvergenceModel2022} has proved that BPG method is convergent algorithm under some assumptions of problems,
and this algorithm can be accelerated according to \Citeauthor{hanzelyAcceleratedBregmanProximal2021}. It is easy to find that we can update 
step sizes by linear search methods in this algorithm, whilst step sizes are obtained by error estimation in RKF method, which costs a lot of time.

$\bm{Notes:}$ Next, we will discuss the relation between optimal methods and ODE methods. Here is an optimal problem: 
\begin{equation*}
    \mathop{min}_{\left(x_1,x_2,\cdot,x_N\right)} E\left(x_1,x_2,\cdots,x_N\right)=f\left(x_1,x_2,\cdots,x_N\right) + g\left(x_1,x_2,\cdots,x_N\right)
\end{equation*}
where $f=\frac{1-d}{2}E\in C^2$  and $g=\frac{1+d}{2}E$ are proper continuous. $d$ is a wieght factor and $-1\leq d\leq 1$. 
It is easy to use BPG method in this optimization problem.

In this problem, it is easy to know:
\begin{equation*}
\begin{aligned}
     f\left(x_1,x_2,\cdots,x_N\right)=\frac{1-d}{2}E\left(x_1,x_2,\cdots,x_N\right)
\end{aligned}
\end{equation*}

And $h\left(x_1,x_2,\cdots,x_N\right)=\frac{1}{2}\sum_{i=1}^N x_i^2$. So, given the result of \emph{k} th, it updates $x^{k+1}$ via
\begin{equation}
\begin{aligned}
      \bm{x}^{k+1}=\mathop{argmin}_{\left(x_1,x_2,\cdots,x_N\right)}\left\{\frac{1+d}{2}E\left(x_1,x_2,\cdots,x_N\right)+\frac{1-d}{2}\sum_{i=1}^N\left(x_i-x_i^k\right)\frac{\partial E}{\partial x_i}^k+\frac{1}{2\alpha^k}\sum_{i=1}^N\left(x_i-x_i^k\right)^2\right\}
      \label{14}
\end{aligned}
\end{equation}
where $\frac{\partial E}{\partial x_i}^k=\frac{\partial E}{\partial x_i}\left(x_1^k,x_2^k,\cdots,x_N^k\right)$. This subproblem can be solved by gradient descent method.
It is not difficult to find that this BPG method is a forward Euler method when $d=-1$ and is a backward Euler method when $d=1$. To some extent, 
we can say that this method is nearly equivalent to the implicit time method \cite{peterffyEfficientImplicitTime2020} as following. When we solve ODE problem $\dot{\bm{x}}=F\left(\bm{x}\right)$, the iteration reads as
\begin{equation}
    \bm{x}^{k+1}=\bm{x}^k+\frac{h}{2}\left(\left(1-d\right)F\left(\bm{x}^k\right)+\left(1+d\right)F\left(\bm{x}^{k+1}\right)\right)
    \label{15}
\end{equation}
where \emph{h} is the step size and $\emph{d}\in \left[-1,1\right]$ is a wieght factor. If we have $\nabla E = -F$ and solve optimal problem \ref{14} by first order optimality conditions, the BPG method
are similar to equation \ref{15} except the step sizes which we use. As the former discussion, if $d=1$, \ref{15} would be the backward Euler method. And if let \emph{d} be -1, this method also is the forward 
Euler method just as the BPG method.

By this way, we can say that an approach from ODE problems can be represented by a certain optimization method and what we need to do 
is advancing an appropriate optimal problem for the ODE system. 

\subsection{The FRCG Method}
As we all know, the linear conjugate gradient method is an important method for solving linear systems with positive coefficient matrices.
What's more, nonlinear variants of the conjugate gradient methods are well studied and it is proved to be quite successful in practice \cite{nocedalNumericalOptimization2006}.
So, we introduce the nonlinear CG method which is called FRCG method here \cite{daiNonlinearConjugateGradient1999}, and it follows the 
algoirthm 

In this method, we get $\alpha^k$ according to the same step size estimation as the BPG method. If we choose \emph{f} to be a strongely convex quadratic and $\alpha^k$ to be the exact minimizer, this algorithm reduces to the linear conjugate gradient method.
In practice, in order to solve problems effectively, we require the step size $\alpha^k$ to satisfy the \emph{strong Wolfe conditions}:
\begin{equation*}
\begin{aligned}
    f\left(x^k+\alpha^k p^k\right)&\leq f\left(x^k\right)+c_1\alpha^k \nabla f^T_k p^k\\
    \vert \nabla f\left(x^k+\alpha^k p^k\right)^Tp^k\vert&\leq -c_2\nabla f^T_k p^k\\
\end{aligned}
\end{equation*}
where $0<c_1<c_2<0.5$, and this property will ensure that all directions $p^k$ are descent directions for the function \emph{f}. 

It is well-known that FRCG method is efficient for some problems and faster than gradient descent method, but the convergence of conjugate gradient 
method for non-linear problems is not proved untill now. To our surprise, this method show some advantages compared to RKF method and BPG method according
to following numrical experiments.

\section{Our New form}
In this chapter, we will introduce the non-singular stress field, which makes sure that the energy and forces can be finite and computational. 
And then, the classical dislocation dynamics would be changed into optimal problems and some efficient methods, such as BPG method and FRCG method, which would be used to accelerate 
simulations. By this way, we will show that it is not difficult to apply optimal approachs in 2D dislocation dynamics.

\subsection{New Optimization Problem}
Here, we will change 2D DD to energy optimal problems for three cases from Section 2 and introduce the non-singular stress field to make sure 
that energy problems are computational.
\subsubsection{Case 1}
Fristly, for Case 1, we can get the energy from equation \ref{1} and \ref{2} with which we can calculate the work done by dislocation interacting force.
The energy minimization formula of Case 1 is \cite{hirthTheoryDislocations1992} 
\begin{equation}
    \mathop{min}_{\left(x_1,x_2,\cdot,x_N\right)} E\left(x_1,x_2,\cdots,x_N\right)=\sum_{i=1}^N\sum_{j=i+1}^N W_{ij}-\sum_{i=1}^N \sigma_{12}b_{x,i}x_i
\end{equation}
where $\sigma_{12}$ is applied stress for the system. And $W_{ij}$ is the interacting energy between dislocation \emph{i} and \emph{j}, 
\begin{equation}
\begin{aligned}
        W_{ij}&=-\frac{\mu\left(\bm{b_i}\cdot\bm{\xi}\right)\left(\bm{b_j}\cdot\bm{\xi}\right)}{2\pi}ln\frac{\rho}{R_a}-\frac{\mu}{2\pi\left(1-\nu\right)}\left[\left(\bm{b_i}\times\bm{\xi}\right)\cdot\left(\bm{b_j}\times\bm{\xi}\right)\right]ln\frac{\rho}{R_a}\\
        &-\frac{\mu}{2\pi\left(1-\nu\right)\rho^2}\left[\left(\bm{b_i}\times\bm{\xi}\right)\cdot\bm{R}\right]\left[\left(\bm{b_j}\times\bm{\xi}\right)\cdot\bm{R}\right]
\end{aligned}
\end{equation}
where $\rho = \sqrt{\left(x_i-x_j\right)^2+\left(y_i-y_j\right)^2}$, $R_a$ is a very big constant to help us compute elastic energy. It is easy to know that 
this energy is consistent with the stress field from equation \ref{1}, and this means that the force from equation \ref{2} is the negative derivative of the energy.
So, this energy minimization problem is equivalent to the dynamic equation \ref{2}.

Obviously, this energy minimization is not easy to solve because the energy may be infinite. It is necessary to remove the singularity of the classic
continuum dislocation theory. According to \Citeauthor{caiNonsingularContinuumTheory2006}, for an edge dislocation system without periodic conditions, the 
non-singular solution can be obtained by introducing a Burger vector density function, and is given by 
\begin{equation}
\begin{aligned}
    \sigma_{xx}^{ns}&=-\frac{\mu }{2\pi \left(1-\nu\right)} \left[\frac{b_xy\left(3x^2+y^2+3a^2 \right)}{\left( x^2+y^2+a^2 \right) ^2}+\frac{b_yx\left(y^2-x^2-a^2 \right)}{\left( x^2+y^2+a^2 \right) ^2}\right]\\
    \sigma_{yy}^{ns}&=\frac{\mu }{2\pi \left(1-\nu\right)}\left[ \frac{b_xy\left(x^2-y^2-a^2 \right)}{\left( x^2+y^2+a^2 \right) ^2}+\frac{b_yx\left(x^2+3y^2+3a^2 \right)}{\left( x^2+y^2+a^2 \right) ^2}\right]\\
    \sigma_{xy}^{ns}&=\frac{\mu }{2\pi \left(1-\nu\right)} \left[\frac{b_xx\left(x^2+a^2-y^2 \right)}{\left( x^2+y^2+a^2 \right) ^2}+\frac{b_yy\left(x^2-a^2-y^2 \right)}{\left( x^2+y^2+a^2 \right) ^2}\right]\\
    \sigma_{xz}^{ns}&=\sigma_{yz}^{ns}=0
    \label{16}
\end{aligned}
\end{equation}
As $a\rightarrow0$, the classical singular solution for the stress field about an infinite straight edge dislocation is recovered. 

According to equation \ref{16}, the new non-singular elastic energy would be a new form

Here should be the non-singular energy formula.

where $\rho_a=\sqrt{\left(x_i-x_j\right)^2+\left(y_i-y_j\right)^2+a^2}$, and $a$ is an arbitrary constant(dislocation core width). $R_a$ is a constant to 
help to calculate the energy.

It is clear that the elastic energy from \ref{16} is finite and bounded below if we assump that dislocations are in a very big box. This non-singular stress and energy make it possible
to solve 2D model by minimizing energy of disloation systems. We can rewrite the original problem \ref{2} into a new optimal problem, which is
\begin{equation}
    \mathop{min}_{\left(x_1,x_2,\cdot,x_N\right)} E\left(x_1,x_2,\cdots,x_N\right)=\sum_{i=1}^N\sum_{j=i+1}^N W_{ij}^{NS}-\sum_{i=1}^N \sigma_{12}b_{x,i}x_i
    \label{18}
\end{equation}

When we use gradient descent method to solve problem \ref{18}, it is equivalent to the Euler method for original form \ref{2}, because for each dislocation \emph{i} there would be
\begin{equation*}
    \frac{\partial E}{\partial x_i}=-\frac{d x_i}{d t}
\end{equation*}
The self-consistency is maintained, which means that the energy minimization form of Case 1 is the same as the dynamic equation of it and we can solve 2D DD by minimizing the interacting energy.

\subsubsection{Case 2}
In Case 2, according to \Citeauthor{kuykendallConditionalConvergenceTwodimensional2013},  we can get infinite summations of the stress fields, and these infinite sums are conveniently evaluated by 
application of the Residue Theorems of complex function theory. The results reads
\begin{equation}
\begin{aligned}
        \sigma_{xy}^{x}&=\frac{\mu}{2\left(1-\nu\right)L}\left[\frac{b_x\left(C_y-c_x-y_p LS_y\right)}{\left(C_y-c_x\right)^2}-\frac{b_y y_p L\left(C_yc_x-1\right)}{\left(C_y-c_x\right)^2}\right]\\
        \sigma_{yy}^{x}&=\frac{\mu }{2\left(1-\nu\right)L}\left[-\frac{b_x y_p L\left(C_yc_x-1\right)}{\left(C_y-c_x\right)^2}+\frac{b_y\left(C_y-c_x+y_p LS_y\right)}{\left(C_y-c_x\right)^2}\right]\\
        \sigma_{xx}^{x}&=\frac{\mu }{2\left(1-\nu\right)L}\left[\frac{b_x\left(y_p L\left(C_yc_x-1\right)-S_y\left(C_y-c_x\right)\right)}{\left(C_y-c_x\right)^2}\right.\\
        &\left.-\frac{b_y \left(C_y-c_x-y_p L S_y\right)}{\left(C_y-c_x\right)^2}\right]\\
        \sigma_{xz}^{x}&=\sigma_{yz}^{x}=0\\
        C_y&=\cosh{y_p}, c_x=\cos{x_p}\\
        S_{y}&=\sinh{y_p}, s_x=\sin{x_p}\\
        \label{19}
\end{aligned}
\end{equation}
where $x_p=\frac{2\pi\left(x_i-x_j\right)}{L}$ and $y_p=\frac{2\pi\left(y_i-y_j\right)}{L}$, and \emph{L} is the length of the simulation box among \emph{x} axis. 
From this stress field, the interacting energy in Case 2 is 

Here should be the non-singular energy formula.
It is not difficult to check that the force obtained by Peach-Koehler formula and the stress field \ref{19} is the negative derivative of this energy.

We can obtain the energy optimization problem for Case 2 in following formula:
\begin{equation}
    \mathop{min}_{\left(x_1,x_2,\cdot,x_N\right)} E\left(x_1,x_2,\cdots,x_N\right)=\sum_{i=1}^N\sum_{j=i+1}^N W_{ij}^x-\sum_{i=1}^N \sigma_{12}b_{x.i}x_i
\end{equation}
This optimal problem is equivalent to equation \ref{4}, because we have
\begin{equation*}
    \frac{\partial E}{\partial x_i}=-\frac{d x_i}{d t}=-\frac{1}{B}\left(\sum_{j \neq i}f_{x,ij}+\sigma_{12}b_{x,i}\right)
\end{equation*}
where $f_{x,ij}$ is defined by \ref{3}. Similarly, the interacting energy from this way might be infinite, and what we need to do is to remove 
the singularity of stress field.

In order to remove the singularity of original field
, we rewrite the energy and the new non-singular elastic energy is\cite{caiNonsingularContinuumTheory2006} 

where $a$ is a constant, and if $a\rightarrow0$, the non-singular stress field and energy will be equal to the classical singular solution.

Following the same idea as Case 1, we can obtain the new problem under periodical condition in \emph{x} axis, it is
\begin{equation}
    \mathop{min}_{\left(x_1,x_2,\cdot,x_N\right)} E\left(x_1,x_2,\cdots,x_N\right)=\sum_{i=1}^N\sum_{j=i+1}^N W_{ij}^{x,NS}-\sum_{i=1}^N \sigma_{12}b_{x.i}x_i
    \label{22}
\end{equation}
Solving this problem is to solve the original form of Case 2, too. For the systems which are infinite in \emph{x} axis and periodic in \emph{y} axis, 
the same treatment can be used and definitely we can deal with those systems easily.

\subsubsection{Case 3}
In Case 3, the new form is similar to the first problem, and we just apply dislocation systems with periodical conditions by a certain truncation scheme. It is easy to consider the periodic conditions
by truncating infinite sums with the square scheme, and the minimization problem for \ref{6} is
\begin{equation}
    \mathop{min}_{\left(\bm{x}_1,\bm{x}_2,\cdot,\bm{x}_N\right)} E\left(\bm{x}_1,\bm{x}_2,\cdots,\bm{x}_N\right)=\sum_{i=1}^N\sum_{j\neq i}^N W_{ij}^{p}-\sum_{i=1}^N \sigma_{12}b_{x,i}x_i-\sum_{i=1}^N \sigma_{12}b_{y,i}y_i
    \label{23}
\end{equation}
Where $W_{ij}^{p}$ can be obtained by equation \ref{7} and elastic energy theory easily and read 
\begin{equation}
\begin{aligned}
    W_{ij}^{p}&=\sum_{m,n=-N_c}^{N_c} W^{cell}\left(\bm{x}_i,\bm{x}_j,m,n\right)\\
    W^{cell}\left(\bm{x}_i,\bm{x}_j,m,n\right)&=W\left(x_i-x_j-mL_x,y_i-y_j-nL_y\right)
    \label{24}
\end{aligned}
\end{equation}
By this way, it is easy to get the energy minimization problems for Case 3. And there is the same disadvantage as Case 1, which is that the energy might not be
computable. All we need to do is to replace the classic energy by non-singualr energy from equation \ref{16}. The new optimization problem is
\begin{equation}
    \mathop{min}_{\left(\bm{x}_1,\bm{x}_2,\cdot,\bm{x}_N\right)} E\left(\bm{x}_1,\bm{x}_2,\cdots,\bm{x}_N\right)=\sum_{i=1}^N\sum_{j\neq i}^N W_{ij}^{p,NS}-\sum_{i=1}^N \sigma_{12}b_{x,i}x_i-\sum_{i=1}^N \sigma_{12}b_{y,i}y_i
    \label{25}
\end{equation}

Where $W_{ij}^{p,NS}$ can be obtained by equation \ref{24} and replace the energy by the non-singular energy between dislocation \emph{i} and \emph{j}:
\begin{equation*}
        W^{cell,NS}_{ij}\left(\bm{x}_i,\bm{x}_j,m,n\right)=W^{NS}\left(x_i-x_j-mL_x,y_i-y_j-nL_y\right)
\end{equation*}

Till now, we change three cases to energy minimization problems, and we just look at 2D dislocation dynamics from a new point of view. The new optimization problems
are equivalent to the original questions which are presented in Section 2. 

\subsection{The New Methods}
Next, we need to solve optimization problems. It is clear that we can consider three minimization problems as one form:
\begin{equation}
    \mathop{min}_{\left(\bm{x}_1,\bm{x}_2,\cdot,\bm{x}_N\right)} E\left(\bm{x}_1,\bm{x}_2,\cdots,\bm{x}_N\right)=f\left(\bm{x}_1,\bm{x}_2,\cdots,\bm{x}_N\right) + g\left(\bm{x}_1,\bm{x}_2,\cdots,\bm{x}_N\right)
    \label{26}
\end{equation}
where $f=E-g\in C^2$ is proper but nonconvex and $g=\frac{\mu}{2\left(1-\nu\right)}\Vert\cdot\Vert^2$ is proper continuous, and convex.
It is easy to use BPG method and FRCG method in this optimization problem.

Next, we will take problem \ref{18} as an example and apply the BPG method to solve it. In this problem, we only focus on movement of dislocations in \emph{x} direction.
It is easy to know:
\begin{equation}
\begin{aligned}
     f\left(x_1,x_2,\cdots,x_N\right)=\sum_{i=1}^N\sum_{j=i+1}^N W_{ij}^{NS}-\sum_{i=1}^N \sigma_{12}x_i-\frac{\mu}{2\left(1-\nu\right)}\sum_{i=1}^N x_i^2
     \label{27}
\end{aligned}
\end{equation}

And $h\left(x_1,x_2,\cdots,x_N\right)=g\left(x_1,x_2,\cdots,x_N\right)=\frac{\mu}{2\left(1-\nu\right)}\sum_{i=1}^N x_i^2$. So, given the result of \emph{k} th, it updates $x^{k+1}$ via
\begin{equation}
\begin{aligned}
      x^{k+1}=\mathop{argmin}_{\left(x_1,x_2,\cdots,x_N\right)}\left\{g\left(x_1,x_2,\cdots,x_N\right)+\sum_{i=1}^N\left(x_i-x_i^k\right)\frac{\partial f}{\partial x_i}^k+\frac{\mu}{2\alpha^k\left(1-\nu\right)}\sum_{i=1}^N\left(x_i-x_i^k\right)^2\right\}
      \label{28}
\end{aligned}
\end{equation}

This subproblem is easy to solve even if $h\left(x_1,x_2,\cdots,x_N\right)=\frac{\mu}{2\left(1-\nu\right)}\Vert\cdot\Vert^4$, which can be solved by Newton method easily. So we 
successfully apply BPG method in dislocation dynamics which have been rewritten to a energy minimization problem. In the same way, it is easy to introduce FRCG method or other optimal methods 
to those problems of dislocation dynamics. The convergence analysis of BPG method is given in Appendix \ref{appendix}. In this paper, there will be three approaches-RKF method, BPG method 
and FRCG method to solve 2D dislocation dynamics and we will compare those methods later. And following this idea, it is easy to extend problems to dislocation systems with multi slip planes. 

\section{Numerical Results}
Here are some numerical results which can show the advantages of new methods, including BPG method and FRCG method. A version of the Matlab code is available on github under
MIT License: \href{https://github.com/Hyt1215/2dDDD.git}{https://github.com/Hyt1215/2dDDD.git}. And all following results are obtained by this code in Windows, Interl(R)
Core i5-9300H of 2.40GHz and Cruial DDR4 2666MHz of 8 GB.

At first, before solving problems, the computational units are chosen so that all the distances are measured in units of the Burgers vector length $b$, 
time is measured in units of $\frac{2 B\pi\left(1-\nu\right)}{\mu b}$, stress in units of $\frac{\mu}{2\pi\left(1-\nu\right)}$ and energy in units of $\frac{\mu}{4\pi\left(1-\nu\right)}$.
These systems are given time to relax with $\sigma_{12}=0$, and in this period the changes of energy and the errors between current states and stable states are shown in 
following results. After relaxation, the loading procedure is started and the simulations for the regression of the stress response used the quasistatic 
stress ramp, which tries to mimic slow compression experiments. During the ramp, the external stress is increased with a certain rate $\Dot{\sigma}_{12}=3\times 10^{-4}$ 
and the stress responses are shown in the following figures.

\subsection{Results of Case 1}
Facing dislocation systems without periodic conditions, we randomly create a system where \emph{N=100}, and then we can apply three methods to solve this problem.

\begin{figure}[H]
    \centering
    \subfigure[Energy]{
    \label{fig1a}
    \includegraphics[width=0.3\textwidth]{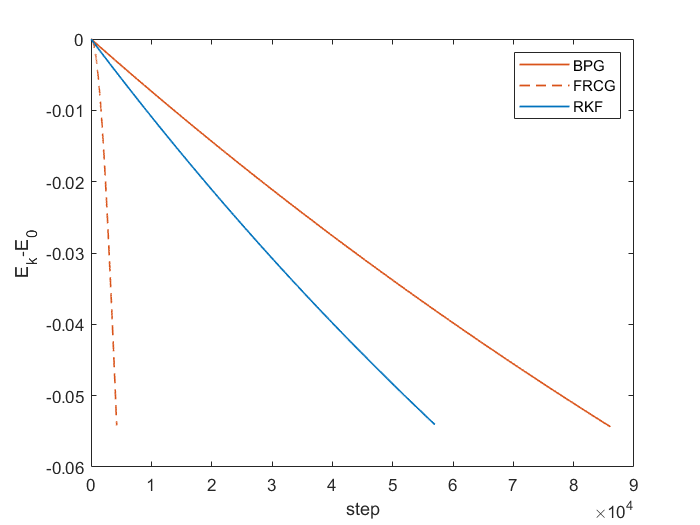}}
    \subfigure[Error]{
    \label{fig1b}
    \includegraphics[width=0.3\textwidth]{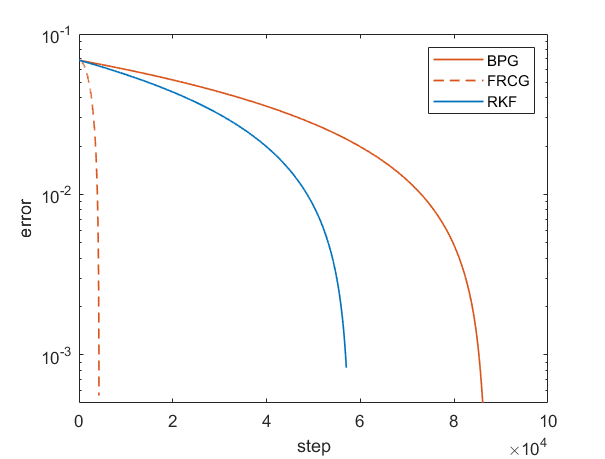}}
    \subfigure[Mean Velocity]{
    \label{fig1c}
    \includegraphics[width=0.3\textwidth]{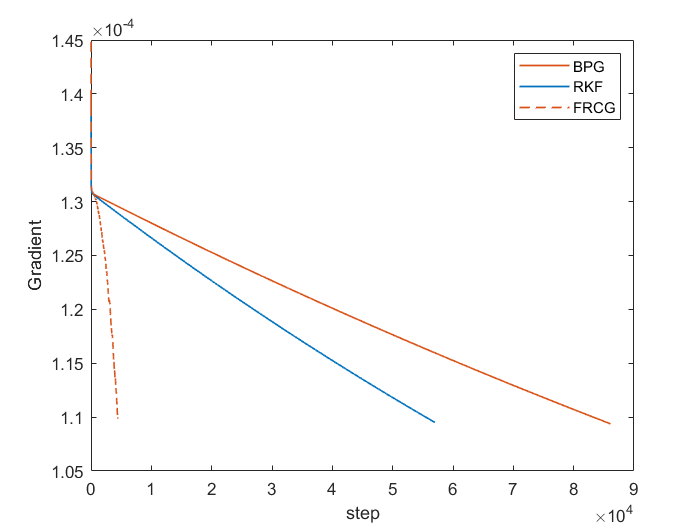}}
    \caption{Results of test 1}
    \label{fig1}
\end{figure}

First, we can focus on the energy relaxing process, and the following results show that the total interacting energy of the system decreased and the converge rates 
of three ways are different. And in Figure \ref*{fig1b}, $error = \frac{1}{N}\sum_{i=1}^N\sqrt{\left(x_i^k-x_i^*\right)^2+\left(y_i^k-y_i^*\right)^2}$, where $\left(x_i^*
,y_i^*\right)$ is the position of dislocation \emph{i} when the system is stable.
In figure \ref{fig1}, it is easy to find that the FRCG method is faster than other methods. Because we use the mean velocity as a condition to exit the 
simulation, the simulation calculated by FRCG method is finished in only 5000 steps and we can find this in figure \ref{fig1c}. However, by BPG method and RKF method, the loops need over ten times 
as many as steps. In this case, the BPG method is slower than the RKF method and this may result from the linear search algorithm for step sizes which limit
the step size and slow down the calculation. In figure \ref{fig1b}, it seems that the rates of convergence of three method are both superlinear but the reason for these curves 
might be the stable state which is used to get errors but only can be obtained by real computational processes. In fact, we can think that the rates of convergence are still sublinear from figure \ref{fig1c},
and this phenomenon is more obvious in Case 3, from figure \ref{fig5c}.

\begin{figure}[H]
    \centering
    \includegraphics[width=0.5\textwidth]{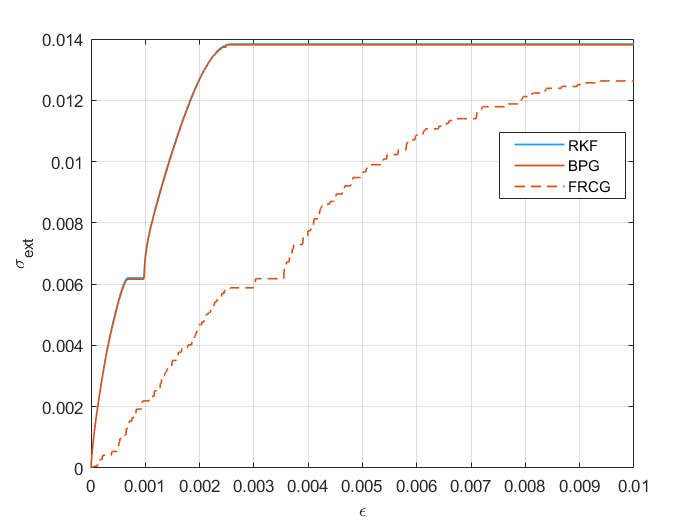}
    \caption{Strain vs Stress for test 1}
    \label{fig2}
\end{figure}

Using different methods, we can predict strain-stress curves as figure \ref{fig2}. In this experiment, it is easy to find that the results from the BPG method and 
the RKF method are almost the same and the reason may be that the BPG method \ref{27} is very similar to RKF method except the step sizes. So, if step sizes of two 
methods are similar, the results from two methods would be close. However, the result obtained by FRCG method is obviously lower than others and this might mean that
the perdiction from FRCG method may be more conservative. This might mean that FRCG method cannot be used in predicting material properties. However, when it comes to 
finding a stationary state of dislocation systems, the FRCG method still is very useful and effective.

\subsection{Results of Case 2}
When it comes to dislocation systems which are applied with periodic condition in \emph{x} direction, similar results can be obtained focusing on a system(\emph{N=100}).
In this case, burger vectors of all dislocations are still $\bm{b}=\left(\pm 1,0,0\right)$.

\begin{figure}[H]
    \centering
    \subfigure[Energy]{
    \label{fig3a}
    \includegraphics[width=0.3\textwidth]{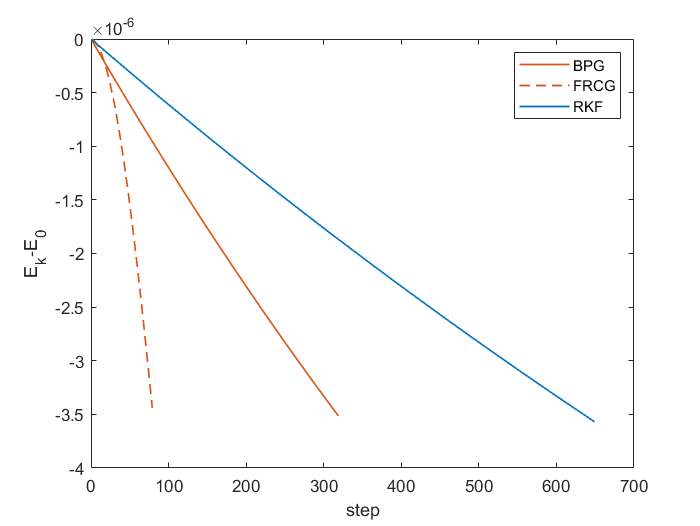}}
    \subfigure[Error]{
    \label{fig3b}
    \includegraphics[width=0.3\textwidth]{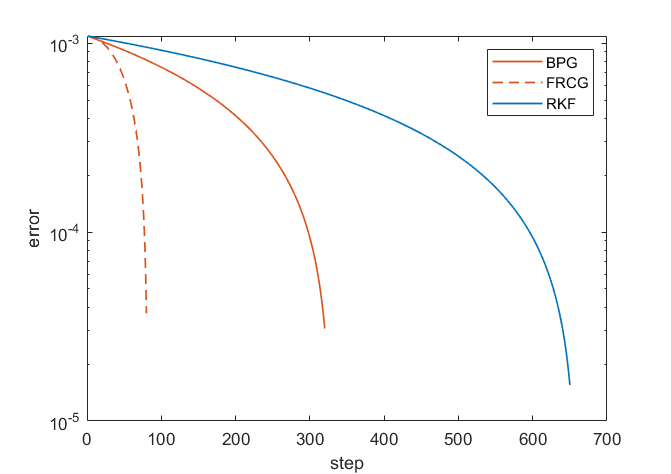}}
    \subfigure[Mean Velocity]{
    \label{fig3c}
    \includegraphics[width=0.3\textwidth]{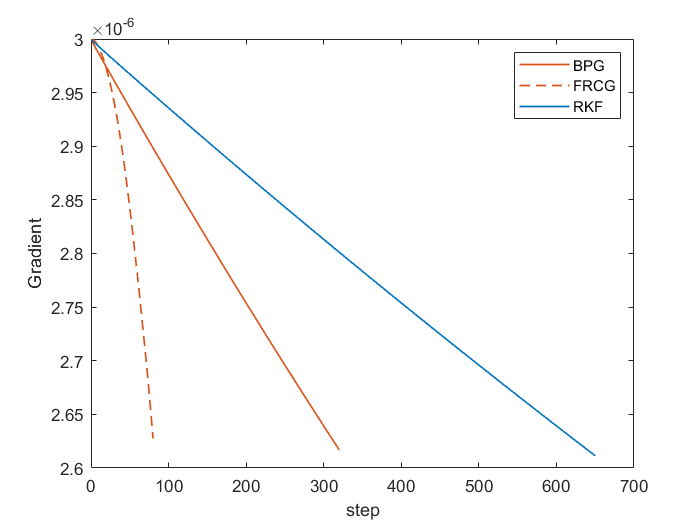}}
    \caption{Results of test 2}
    \label{fig3}
\end{figure}
In figure \ref{fig3}, we can know that the FRCG method still is the fastest method to calculate the simulation as Case 1. To our surprise, in Case 2
the BPG method solves the problem faster than the RKF method and the descent rates of the energy, error and mean velocity of dislocation system
in BPG method are higher than those in RKF method. So, Case 2 shows some advantages of the BPG method compared to the RKF method. In some experiments,
the BPG method may be slower than the FRCG method but faster than the RKF method.

\begin{figure}[H]
    \centering
    \includegraphics[width=0.5\textwidth]{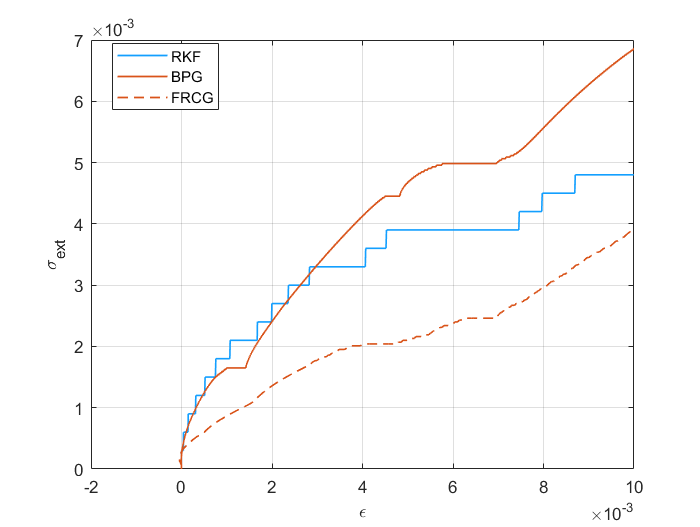}
    \caption{Strain vs Stress for test 2}
    \label{fig4}
\end{figure}

Predicting the curve of strain vs stress in figure \ref{fig4}. In this figure, we can find that the result from FRCG method definitely is lower than that 
from other methods. Out of our expection, the result calculated by BPG method is higher than by the RKF method but it is hard to tell which result is 
closer to the reality from a mathematical point of view. 

\subsection{Results of Case 3}
Because this case is much more complex than others, we will concentrate on an example with \emph{N=48} for systems where there 
are periodical in \emph{x} and \emph{y} axis.

\begin{figure}[H]
    \centering
    \subfigure[Energy]{
    \label{fig5a}
    \includegraphics[width=0.3\textwidth]{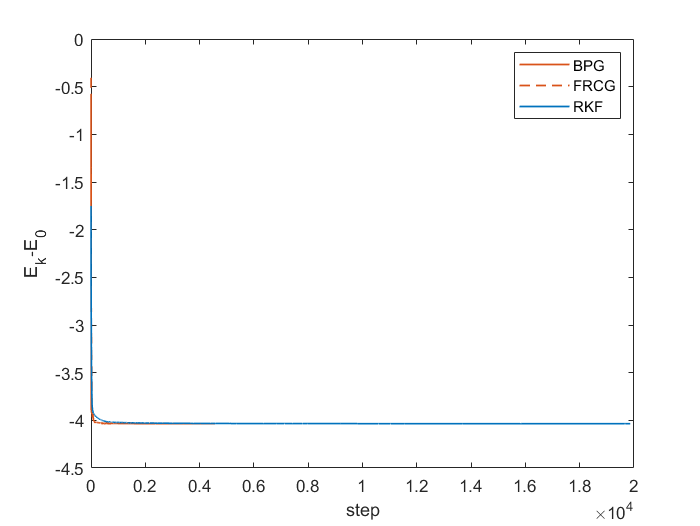}}
    \subfigure[Error]{
    \label{fig5b}
    \includegraphics[width=0.3\textwidth]{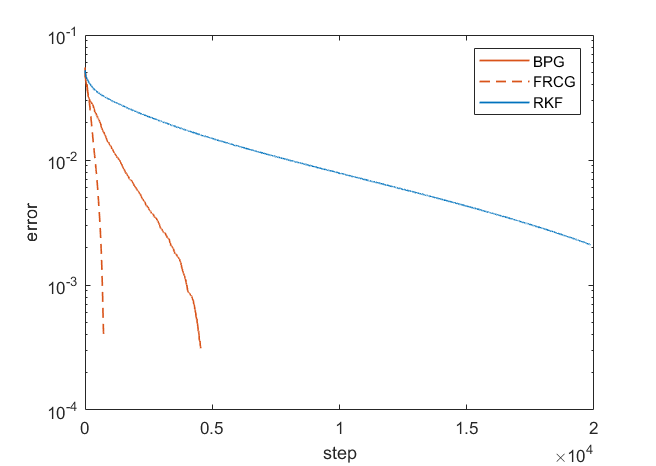}}
    \subfigure[Mean Velocity]{
    \label{fig5c}
    \includegraphics[width=0.3\textwidth]{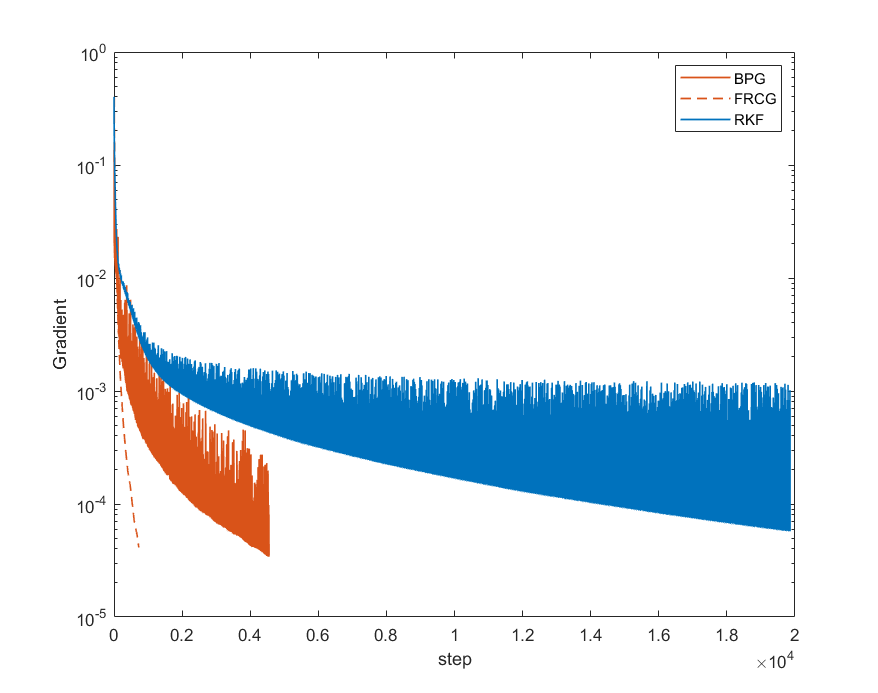}}
    \caption{Results of test 3}
    \label{fig5}
\end{figure}
This case might be closer to the real world than other cases and this means that Case 3 is more important. Here, we can make a conclusion that 
the FRCG method obviously is faster than other methods we studied in all cases. As Case 2, the BPG method is more efficient than the RKF method.
However, according to the mean velocity (in figure \ref{fig5c}), the rates of convergence of all methods are sublinear and the simulation might require
$O\left(\frac{1}{\epsilon^2}\right)$ iterations when we find the first stationary point just as the gradient descent method for nonconvex problems \cite{jainGradientMethodsNonconvex2019}. 

Meanwhile, we can get strain for this systems when we apply external stress. It is easy to find that in Case 3, the result obtained from the RKF method
is higher than other methods. The results from the BPG method and the FRCG method are more similar. 
\begin{figure}[H]
    \centering
    \includegraphics[width=0.5\textwidth]{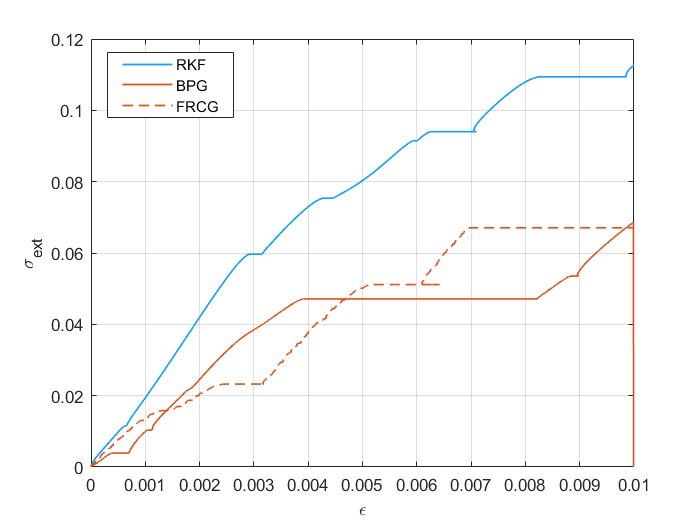}
    \caption{Strain vs Stress for test 3}
    \label{fig6}
\end{figure}

In summary, it is clear that FRCG method is the most efficient method among three methods. And BPG method is a little
bit better than RKF method, especially when it comes to complex systems such as Case 3. From the strain-stress curves, we can know that 
the prediction for stress by the FRCG method usually is lower than by the RKF method. So, we may not 
use the FRCG method to simulate material behaviours but we can use it to find a stable state for dislocation systems, which also is 
a difficult and meaningful problem.

\section{Conclusions}
Dislocation dynamics play a central role in simulating material behaviours under deformation nowadays. Although the 2D model is easier to compute 
than the 3D model, the 2D model still costs much time especially for a large dislocation system. At the same time, most pervious research focuses on
solving 2D dislocation dynamics as ODE systems, and a few properties like energy are not considered in simulations. In this work, we rewrote the 2D model
into a energy minimization problems and applied several optimal methods to solve our problems. 

In classic continuum dislocation theory, the interacting energy and stress field may be infinite and this means that the energy minimization from the classical
theory is not solvable. So, the non-singular stress field theory for dislocations is introduced into our problems and based on this our energy optimal problems
can be solved by the BPG method and the FRCG method. In some dislocation systems, especially in large systems, the BPG method may be just a little bit better than the RKF method which is traditional approach for dislocation
dynamics. The FRCG method is definitely the fastest method in our discussion, but the strains predicted by it are smaller than those obtained by the 
RKF method. So, the FRCG method may not be applicable to simulate dislocation dynamic behaviours but can be useful when it comes to finding stable states for dislocation systems.

\begin{appendices}
    \section{Appendix: Convergence of BPG method}\label{appendix}
    Focusing on problem \ref{26}, we will analyze the convergence of the BPG method. More detail about proof in \Citeauthor{jiangEfficientNumericalMethods2020}.

    \textit{Assumption 1} $E$ is bounded below, and for any $ x^0 \in domE$, 
    the sublevel set $\mathcal{M}\left(x^0\right):=\left\{x\vert E\left(x\right) \leq E\left(x^0\right)\right\}$ is compact.

    When the distance between two dislocations is large enough, we believe that there is no interaction between them, and the interaction energy does not exist. Then, we can set 
    the interaction energy to $+\infty$, as $\rho_{ij} > R_a$. Then dislocations always move in a bounded area. It is easy to say \textit{Assumption 1} holds.

    Let $\mathcal{N}=\left\{x\vert E\left(x\right) < \infty \right\}$. From equation \ref{27}, it is easy to know that there exists C which satisfies $\Vert \nabla ^2 f\left(x\right)\Vert< C$ in $\mathcal{N}$. 
    So $\nabla f$ is Lipschitz smoothness: $\vert \nabla f\left(x\right)-\nabla f\left(y\right) \vert \leq C \Vert x-y\Vert$.

    And we can get
    \begin{equation*}
        f\left(x\right)-f\left(y\right)-\left<\nabla f\left(y\right),x-y\right> \leq \frac{C}{2}\Vert x-y\Vert ^2
    \end{equation*}
    where $C$ is a constant. As mentioned before, $D_h\left(x,y\right)=\frac{\Vert x-y\Vert ^2}{2}$, so
    \begin{equation}
        f\left(x\right)-f\left(y\right)-\left<\nabla f\left(y\right),x-y\right> \leq C D_h\left(x,y\right)
        \label{equ1}
    \end{equation}

    \textit{Lemma 1} Let $\alpha > 0$. If
    \begin{equation}
        z=\mathop{\arg\min}\limits_{x}\left\{g\left(x\right)+\left<x-y,\nabla f\left(y\right)\right>+\frac{1}{\alpha}D_h\left(x,y\right)\right\}
        \label{equ2}
    \end{equation}
    then there exists some $\sigma >0$ such that
    \begin{equation}
        E\left(y\right)-E\left(z\right) \geq \sigma \Vert z-y \Vert ^2
        \label{equ3}
    \end{equation}
    \textit{proof}. According to the optimal condition, we have 
    \begin{equation*}
    \begin{aligned}
        E\left(y\right)&=f\left(y\right)+g\left(y\right)=\left[f\left(y\right)+\left<\nabla f\left(y\right),x-y\right>+\frac{1}{\alpha}D_h\left(x,y\right)+g\left(x\right)\right]_{x=y}\\
        &\geq f\left(y\right)+\left<\nabla f\left(y\right),z-y\right>+\frac{1}{\alpha}D_h\left(z,y\right)+g\left(z\right)\\
        &\geq f\left(z\right)-CD_h\left(z,y\right)+\frac{1}{\alpha}D_h\left(z,y\right)+g\left(z\right)\\
        &=E\left(z\right)+\left(\frac{1}{\alpha}-C\right)D_h\left(z,y\right)
    \end{aligned}
    \end{equation*}

    And then we can get
    \begin{equation*}
        E\left(y\right)-E\left(z\right) \geq \left(\frac{1}{\alpha}-C\right) \Vert z-y \Vert ^2
    \end{equation*}
    Let $\sigma=\frac{1}{\alpha}-C$, Equation \ref{equ3} is proved.

    \textit{remark 1}. \textit{Assumption 1} hold. Let $\left\{x^k\right\}$ be the sequence generated by BPG method. Then $\left\{x^k\right\} \subset \mathcal{M}\left(x^0\right)$ and
    \begin{equation}
        E\left(x^k\right)-E\left(x^{k+1}\right) \geq c_0\Vert x^k-x^{k+1}\Vert^2
        \label{equ4}
    \end{equation}
    where $c_0 > 0$ is a constant. The proof is a straightforward result of \textit{Lemma 1}.

    \textit{Lemma 2}(bounded by the gradient) Let $\left\{x^k\right\}$ be the sequence generated by BPG method. Then, there exists $c_1$ such that
    \begin{equation}
        \Vert \nabla E\left(x^{k+1}\right)\Vert \leq c_1\Vert x^k-x^{k+1}\Vert
        \label{equ5}
    \end{equation}

    \textit{proof}. By the first-order optimality condition of Equation \ref{28}, we get
    \begin{equation*}
    \begin{aligned}
        0 &= \nabla f\left(x^k\right)+\frac{1}{\alpha_k}\left(\nabla h\left(x^{k+1}\right)-\nabla h\left(x^k\right)\right)+\nabla g\left(x^{k+1}\right) \\
        \iff & -f\left(x^k\right)-\frac{1}{\alpha_k}\left(\nabla h\left(x^{k+1}\right)-\nabla h\left(x^k\right)\right) = \nabla g\left(x^{k+1}\right)
    \end{aligned}
    \end{equation*}
    Since $f \in C^2$, we know that
    \begin{equation*}
        \nabla E\left(x\right)=\nabla f\left(x\right)+\nabla g\left(x\right)
    \end{equation*}
    Then we have 
    \begin{equation*}
    \begin{aligned}
        \Vert \nabla E\left(x^{k+1}\right)\Vert&= \Vert \nabla f\left(x^{k+1}\right)+\nabla g\left(x^{k+1}\right)\Vert \\
        &= \Vert \nabla f\left(x^{k+1}\right)-\nabla f\left(x^k\right)-\frac{1}{\alpha_k}\left(\nabla h\left(x^{k+1}\right)-\nabla h\left(x^k\right)\right)\Vert \\
        &\leq \Vert \nabla f\left(x^{k+1}\right)-\nabla f\left(x^k\right) \Vert +\frac{1}{\alpha_k}\Vert\nabla h\left(x^{k+1}\right)-\nabla h\left(x^k\right)\Vert \\
        &\leq \left(C+\frac{1}{\alpha_k}\right)\Vert x^{k+1}-x^k\Vert \\
        &\leq c_1\Vert x^{k+1}-x^k\Vert
    \end{aligned}
    \end{equation*}
    where the third inequality is right because $\nabla f$ is Lipschitz smoothness. The \textit{Lemma 2} has been proved.

    \textit{Theorem 1} Let $\left\{x^k\right\}$ be the sequence generated by the BPG method. Then, for any limit point $x^*$ of $\left\{x^k\right\}$, we have $\nabla E\left(x^*\right)=0$.

    \textit{proof}. We know $\left\{x^k\right\} \subset \mathcal{M}\left(x^0\right)$ and thus bounded. Then, the set of limit points of $\left\{x^k\right\}$ is nonempty. For any limit point $x^*$, there is a subsequence $\left\{x^{k_j}\right\}$ such that $\mathop{lim}\limits_{j\rightarrow \infty }x^{k_j}=x^*$. $\left\{E\left(x^k\right)\right\}$ is decreasing sequence, and $E$ is bounded below. Then, there exists some $\overline{E}$ such that $\mathop{lim}\limits_{k\rightarrow \infty}E\left(x^k\right)=\overline{E}$. Moreover, it has 
    \begin{equation}
        E\left(x^0\right)-\overline{E}=\mathop{lim}\limits_{K\rightarrow \infty}\sum_{j=0}^K\left(E\left(x^j\right)-E\left(x^{j+1}\right)\right) \geq c_0\mathop{lim}\limits_{K\rightarrow \infty}\sum_{j=0}^K\Vert x^j-x^{j+1}\Vert^2
        \label{equ6}
    \end{equation}
    and so $\mathop{lim}\limits_{k\rightarrow \infty}\Vert x^k-x^{k-1}\Vert=0$. Together with \textit{Lemma 2}, it implies that 
    \begin{equation}
        \mathop{lim}\limits_{j\rightarrow \infty}\Vert \nabla f\left(x^{k_j}\right)+\nabla g\left(x^{k_j}\right)\Vert =0 \Rightarrow \mathop{lim}\limits_{j\rightarrow \infty}\nabla E\left(x^{k_j}\right)=0
    \end{equation}

    We know that $E$ is continuous function, thus $ \nabla E\left(x^*\right)=0$. \textit{Theorem 1} has been proved.

    Next we will introduce that $E$ is Kurdyka-Lojasiewicz(KL) function, and then the subsequence convergence can be strengthened.

    \textit{Definition 1}(Kurdyka-Lojasiewicz property) Let $\sigma :\mathbb{R}^d \rightarrow \left(-\infty,+\infty\right]$ be proper and lower semicontinuous.

    The function $\sigma$ is said to have KL property at $\overline{u}\in dom \partial \sigma:=\left\{u\in\mathbb{R}^d:\partial\sigma \neq \emptyset\right\}$, 
    if there exist $\eta\in \left(0,+\infty\right]$, a neighborhood $U$ of $\overline{u}$ and a function $\phi \in \Phi_{\eta}$, such that
    \begin{equation*}
        \forall u \in U\cap \left\{u\in \mathbb{R}^d:\sigma\left(\overline{u}\right) <\sigma\left(u\right)<\sigma\left(\overline{u}\right)+\eta\right\}
    \end{equation*}

    The following inequality holds:
    \begin{equation}
        \phi '\left(\sigma\left(u\right)-\sigma\left(\overline{u}\right)\right)dist\left(\mathbf{0},\partial\sigma\left(u\right)\right)\geq 1
        \label{equ7}
    \end{equation}
    where $\Phi_{\eta}$ is the class of all concave and continuous functions $\phi:\left[0,\eta\right)\rightarrow \mathbb{R}_+$which satisfy the following conditions:
    \begin{enumerate}
        \item $\phi\left(0\right)=0$;
        \item $\phi$ is $C^1$ and continuous at 0;
        \item for all $s\in\left(0,\eta\right):\phi '\left(s\right)>0$
    \end{enumerate}

    And $dist\left(\mathbf{0},\partial \sigma\left(x^{k+1}\right)\right)=inf\left\{\Vert y \Vert:y\in\partial \sigma\left(x^{k+1}\right)\right\}$, \par
    $\partial \sigma\left(x\right)=\left\{u:\sigma\left(y\right)-\sigma\left(x\right)-\left<u,y-x\right> \geq 0,\forall y\in dom\sigma\right\}$.
    If $\sigma$ have KL property at each point of $dom \partial\sigma$, $\sigma$ is called a KL function. 
    It is known that many functions are KL functions, including the interaction energy of 2D DDD problem without PBCs. 
    And then we get $dist\left(\mathbf{0},\partial E\left(x\right)\right) = \Vert \nabla E\left(x\right)\Vert$, because $E \in C^2$.

    \textit{Theorem 2} Let $\left\{x^k\right\}$ be the sequence generated by the BPG method. Then, there exists a point $x^*\in \mathcal{M}\left(x^0\right)$ such that 
    \begin{equation}
        \mathop{lim}\limits_{k\rightarrow +\infty}x^k = x^*,\nabla E\left(x^*\right)=0
        \label{equ8}
    \end{equation}

    \textit{proof}. Now we need to prove the sequence $\left\{x^k\right\}$ is the Cauchy sequence. Let $S\left(x^0\right)$ be the set of the limiting points of the 
    sequence $\left\{x^k\right\}$. It is easy to know $S\left(x^0\right)$ is nonempty set. From the convergence of $E\left(x^k\right)$, we know that $E\left(x\right)$ 
    is constant on $S\left(x^0\right)$, denoted by $E^*$. If there exists some $k_0$ such that $E\left(x^{k_0}\right)=E^*$, then we have $E\left(x^k\right)=E^*,\forall k\geq k_0$. 
    So we assume that $E\left(x^k\right)>E^*$, in the following proof. Therefore, $\forall \epsilon,\eta >0$, there exists $K>0$, such
     that $dist\left(S\left(x^0\right),x^k\right)\leq \epsilon,E^*<E\left(x^k\right)<E^*+\eta, \forall k\geq K$. According to KL property, we have
    \begin{equation*}
        \phi '\left(E\left(x^k\right)-E^*\right)\Vert \nabla E\left(x^k\right)\Vert\geq1
    \end{equation*}

    Together with \textit{Lemma 2}, it implies that
    \begin{equation}
        \phi '\left(E\left(x^k\right)-E^*\right)\geq \frac{1}{c_1\Vert x^k-x^{k-1}\Vert}
        \label{equ9}
    \end{equation}
    By the convexity of $\phi$, we have
    \begin{equation*}
        \phi \left(E\left(x^k\right)-E^*\right)-\phi \left(E\left(x^{k+1}\right)-E^*\right)\geq \phi '\left(E\left(x^{k+1}\right)-E^*\right)\left(E\left(x^k\right)-E\left(x^{k+1}\right)\right)
    \end{equation*}
    Define $\Delta_{pq}=\phi \left(E\left(x^p\right)-E^*\right)-\phi \left(E\left(x^q\right)-E^*\right)$, we can have for all $k>K$
    \begin{equation}
        \Delta_{k,k+1}\geq\frac{c_0\Vert x^{k+1}-x^k\Vert^2}{c_1\Vert x^{k+1}-x^k\Vert}
        \label{equ10}
    \end{equation}
    Therefore, we have
    Therefore, we have
    \begin{equation}
        \Vert x^{k+1}-x^k\Vert \leq c\Delta_{k,k+1}
        \label{equ11}
    \end{equation}
    where $c=c_1/c_0$. For any $k>K$, summing up \ref{equ11} for $i=K+1,\cdots,k$, we will get
    \begin{equation*}
    \begin{aligned}
        \sum_{i=K+1}^k\Vert x^{i+1}-x^i\Vert &\leq c\sum_{i=K+1}^k\Delta_{i,i+1}\\
        &\leq c\Delta_{K+1,k+1}
    \end{aligned}
    \end{equation*}
    where the last inequality is from the fact that $\Delta_{pq}+\Delta_{qr}=\Delta_{pr},\forall p,q,r\in \mathbb{N}$.

    Then we can easily get
    \begin{equation}
        \sum_{i=K+1}^k\Vert x^{i+1}-x^i\Vert \leq c\left(\phi\left(E\left(x^{K+1}\right)-E^*\right)-\phi\left(E\left(x^k\right)-E^*\right)\right)
    \end{equation}

    Together with $E\left(x^k\right)\rightarrow E^*, k\rightarrow \infty$, it implies that $\sum_{i=K+1}^k\Vert x^{i+1}-x^i\Vert \rightarrow 0$. 
    Then we have $\left\{x^k\right\}$ is the Cauchy sequence. Together with \textit{Theorem 1}, we obtain
    \begin{equation*}
        \mathop{lim}\limits_{k\rightarrow +\infty}x^k = x^*, \nabla E\left(x^*\right)=0
    \end{equation*}
\end{appendices}

\section{Acknowledgements}

\bibliographystyle{unsrtnat}
\bibliography{sample.bib}

\end{document}